\documentclass[prl,twocolumn,showpacs,amsmath,amssymb]{revtex4-1}

\usepackage{graphicx}
\usepackage{amssymb,amsmath}
\usepackage{dcolumn}
\usepackage{bm}
\usepackage{color}

\begin{document}

\title{Black Holes and Wormholes in spinor polariton condensates}

\author{D.D. Solnyshkov}
\affiliation{LASMEA, Nanostructure and Nanophotonics Group, Clermont Universit\'{e}}
\affiliation{Universit\'{e} Blaise Pascal, CNRS, 63177 Aubi\`{e}re Cedex France}

\author{H. Flayac}
\affiliation{LASMEA, Nanostructure and Nanophotonics Group, Clermont Universit\'{e}}
\affiliation{Universit\'{e} Blaise Pascal, CNRS, 63177 Aubi\`{e}re Cedex France}

\author{G. Malpuech}
\affiliation{LASMEA, Nanostructure and Nanophotonics Group, Clermont Universit\'{e}}
\affiliation{Universit\'{e} Blaise Pascal, CNRS, 63177 Aubi\`{e}re Cedex France}

\date{\today}

\begin{abstract}
We propose a new system for the study of event horizons and black holes - a Bose-Einstein condensate of exciton-polaritons. Hawking radiation is observed in numerical experiment. A closed horizon is obtained in 2D. We simulate inter-Universe and intra-Universe wormholes capitalizing on both the spinor nature of polariton condensates and the spin dependence of polariton-polariton interactions.
\end{abstract}

\pacs{71.36.+c,71.35.Lk}
\maketitle

It happens quite often in Physics that striking similarities are found between systems, which from the first glance have absolutely nothing in common. Sometimes, such similarities can be exploited to perform laboratory studies on accessible objects similar to inaccessible ones. One of the most recent examples is that of Klein tunneling in graphene\cite{Klein}: the quasi-particles in a solid-state object obey the same mathematical equations as very high-energy relativistic particles. Astrophysics allows even less laboratory studies than high-energy physics: scientists are restrained to the objects in the Universe proposed by the Nature, and these are studied from very far. Thus, having a desktop version of a black hole (BH) would be even more useful. Once again, the Physics of the small comes to the aid of the Physics of the large. The analogy between the equations describing the excitations of a Bose-Einstein condensate (BEC -- a peculiar state of matter exhibiting quantum properties at macroscopic scales) and the metrics of the curved space-time has been noticed about a decade ago\cite{Garay}. Since then, the scientists have managed to experimentally observe the event horizons in atomic BECs\cite{Lahav}.

However, such atomic condensates are still a bit far from being a convenient laboratory tool, because they require ultra-low temperatures for their formation, and the measurements of the distributions inside the condensates are relatively complex to carry out. Here the solid-state physics comes into play with exciton-polaritons\cite{Microcavities}: hybrid quasi-particles that share the properties of both light and matter, formed in microcavities in the strong coupling regime. On one hand, their excitonic part allows them to interact both with themselves and the surrounding phonon bath, providing efficient relaxation and formation of a BEC \cite{BECPolaritons}. On the other hand, their photonic part attributes them a very small effective mass which allows BEC even at room temperature \cite{RoomTemp}. The finite (and short) lifetime of polaritons turns into an important advantage, simplifying all measurements, because the decay of the condensate means the emission of photons from the cavity, and the distribution of emitted photons gives direct information on the polaritons distribution function, on their dispersion and spatial evolution, that is, the whole wavefunction ${ {{\mathbf{\psi }}\left( {{\mathbf{r}},t} \right)}}$ up to a constant global phase (using interferometry). Finally, the spin structure of polaritons\cite{ReviewSpin} (two allowed spin projections on the growth axis) implies possible new effects due to the vectorial nature of the condensates. These particularities favor polaritons with respect to other systems proposed for the simulation of BHs, including the optical ones based on metamaterials \cite{Muamer,Chen,Greenleaf}.

In this paper, we propose polariton condensates as a convenient tool for the study of sonic BHs, event horizons and Hawking radiation. We begin with a simpler case of a 1D scalar condensate and then consider a 2D system with a closed horizon. Finally, capitalizing on spin, we demonstrate different configurations containing wormholes, including one allowing the signal to be transmitted faster than the speed of sound.

\emph{Event horizon in 1D flow} It has been understood quite a long time ago in hydrodynamics, that an event horizon can appear if the flow speed increases and becomes larger than the speed of sound \cite{Unruh}.  Indeed, the excitations in the flowing medium propagate with the speed of sound (in the linear approximation, that is, long-wavelength limit), and therefore in the laboratory frame they are unable to go against the flow if its speed is too high. A great research effort in this domain has recently culminated with the observation of stimulated Hawking emission in water \cite{Weinfurtner}.

The weak excitations in the BECs (called Bogolons) can also exhibit linear dispersion in the long-wavelength limit, with the speed of sound given by $c_{s}=\sqrt{\alpha n/m}$, where $\alpha$ is the interaction constant (in the Born approximation), $n$ is the density and $m$ is the mass of bosons composing the condensate. The finite lifetime of polaritons provides a natural way to vary the speed of sound. A propagating condensate can be  injected locally by resonant or non-resonant pumping \cite{Wertz2010}. Its density  is bound to decrease with the distance from the pumping spot, and the speed of sound will therefore decrease as well, whereas the propagation speed will remain constant or may even increase, if the condensate is accelerated by some potential ramp (for example, by its own self-interactions). At some point the two speeds become equal, defining the position of the event horizon of the BH.

To describe accurately the polariton condensate, we use a set of standard equations for the photonic $\psi_{ph}(x,y,t)$ and excitonic $\psi_{ex}(x,y,t)$ mean fields coupled via the light matter interaction $\Omega_R=15$ meV, accounting for the two allowed spin projections $\sigma=\pm1$:
\begin{eqnarray}
i\hbar \frac{{\partial \psi _{ph}^\sigma }}{{\partial t}} &=&  - \frac{{{\hbar ^2}}}{{2{m_{ph}}}}\Delta \psi _{ph}^\sigma + \frac{\Omega_R }{2}\psi _{ex}^\sigma \\
\nonumber &-& \frac{{i\hbar }}{{2{\tau _{ph}}}}\psi _{ph}^\sigma  + {P^\sigma } + U(x)\psi _{ph}^{ \sigma } + H_{eff}\psi _{ph}^{ - \sigma }\\
i\hbar \frac{{\partial \psi _{ex}^\sigma }}{{\partial t}} &=&  - \frac{{{\hbar ^2}}}{{2{m_{ex}}}}\Delta \psi _{ex}^\sigma + \frac{\Omega_R }{2}\psi _{ex}^\sigma \\
\nonumber &-& \frac{{i\hbar }}{{2{\tau _{ex}}}}{\psi _{ex}^\sigma} + \left( {{\alpha _1}{{\left| {\psi _{ex}^\sigma } \right|}^2} + {\alpha _2}{{\left| {\psi _{ex}^{ - \sigma }} \right|}^2}} \right)\psi _{ex}^\sigma
\end{eqnarray}
Here $m_{ph}=3.6\times10^{-5}m_0$, $m_{ex}=0.4 m_0$ and $m_0$ are the cavity photon, the quantum well exciton and the free electron masses respectively. The lifetime of the particles are $\tau_{ph}=25$ ps and $\tau_{ex}=300$ ps. $U(x)$ is the potential (e.g. disorder). $H_{eff}(x,y)$ accounts for the possible effective magnetic field coupling the two spin components. The interaction constants $\alpha_1=0.02$ meV$/\mu$m$^2$
and $|\alpha_2|<<|\alpha_1|$
\cite{Renucci2005}
correspond to parallel and anti-parallel spins. Finally, $P^\sigma(x,y,t)$ is a photonic pump tuned in quasi-resonance with the lower polariton branch.

\begin{figure}
  \includegraphics[width=0.45\textwidth,clip]{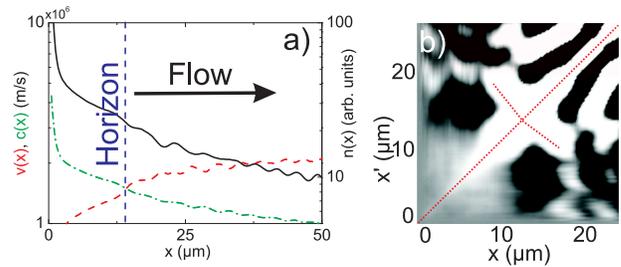}\\
  \caption{(Color online) a) Propagation of a polariton condensate in 1D: density (black solid line), flow speed (red dashed line), sound speed (green dash-dotted line). The horizon is indicated by a blue vertical line. b) Density-density correlation matrix $g^{(2)}(x,x')$ at $t=200~$ps. Red lines are a guide for the eyes, indicating the Hawking tongues (positive correlations) extending from the main diagonal.}
  \label{Fig1}
\end{figure}

First, we consider a 1D case neglecting the spin degree of freedom. The polariton flow is resonantly and locally injected by a pumping laser located close to $x=0$ on the figure \ref{Fig1}(a). The medium is assumed to exhibit small structural disorder mainly due to the etching of the 1D wire cavities. It is modeled by a random series of delta-peaks separated by 1 $\mu$m on average. All results shown in Fig.\ref{Fig1} are averaged over 100 disorder realizations. The Fig.\ref{Fig1}(a) shows the polariton density $n(x)$, the visible decay is mainly due to the finite lifetime of polaritons. The speed of the sound $c(x)$ decreases together with the density. On contrary, the speed of the flow $v(x)$ is increasing because of the self-interactions within the condensate. In the left part, the flow is subsonic and cannot be scattered by the disorder (superfluid). In the right part, the flow is supersonic. The exponential decay is induced not only by the life time but also because of the Anderson like-localization in the disorder. The two regions are separated by a horizon at $v=c$. In the supersonic region no excitation can propagate towards the horizon. The generation of Hawking emission on the horizon is demonstrated in the Fig.\ref{Fig1}(b). Indeed as recently proposed in Refs.\cite{Carusotto2008,CarusottoNJP}, emission of Hawking phonons means correlated density perturbations propagating on both sides of the horizon. Hawking emission can therefore be detected using the density-density correlation matrix ${g^{\left( 2 \right)}}\left( {x,x'} \right) = {{\left\langle {n\left( x \right)n\left( {x'} \right)} \right\rangle } \mathord{\left/
 {\vphantom {{\left\langle {n\left( x \right)n\left( {x'} \right)} \right\rangle } {\left\langle {n\left( x \right)} \right\rangle \left\langle {n\left( {x'} \right)} \right\rangle }}} \right.
 \kern-\nulldelimiterspace} {\left\langle {n\left( x \right)} \right\rangle \left\langle {n\left( {x'} \right)} \right\rangle }}$. This matrix, averaged on disorder, is shown at $t=200~$ps in Fig.\ref{Fig1}(b). Indeed, as expected, the "Hawking tongues", indicating positive correlations, are extending from the horizon position, marked by the red dashed lines. We underline that we do not need to introduce quantum fluctuations to seed Hawking emission thanks to the presence of the disorder potential and to the finite lifetime which broaden the states in momentum and frequency.

\emph{2D black holes} The finite lifetime of polaritons allows to organize persistent flows, as shown in ref. \cite{Lagoudakis2008,DeveaudScience, Wertz2010} for the cases of quasi-resonant or non-resonant pumping. This particular property makes the formation of a closed event horizon in 2D possible. This is much more complicated with atomic condensates, where only 1D configurations have been considered\cite{Lahav}. With polaritons, one needs to pump a region around a large-scale defect in the microcavity mirrors possessing a lower quality factor. 
Polariton-polariton interactions will create a persistent flow converging into the defect region, where the density is always lower due to the shorter lifetime.

\begin{figure}
  \includegraphics[width=0.4\textwidth,clip]{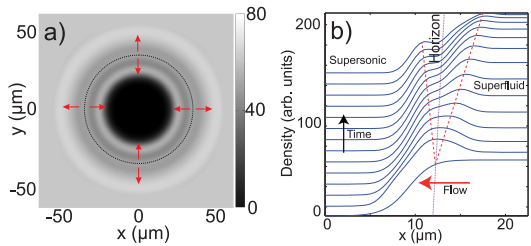}\\
  \caption{(Color online) a) 2D BH around a defect (dark region) in a polariton condensate. Dashed line shows the event horizon, and red arrows point the directions of the propagation of the Hawking radiation on both sides of the horizon. b) Waterfall density plot at different times. Propagating Hawking phonons are marked by the red dotted lines.}
  \label{Fig2}
\end{figure}

Figure \ref{Fig2} shows the results of a realistic 2D simulation with pulsed spatially homogeneous pumping. The photon density $\left|\psi_{ph}\right|^2$ at the time $t=8$ ps is plotted as a function of coordinates (panel a). The defect region with a shorter lifetime is located at the origin ($(x,y)=(0,0)$), while the disorder is neglected. The density inside the defect region decreases faster than outside, and the repulsive interactions make polaritons propagate towards the center of the figure. The event horizon at that time is marked with a dashed line. The Hawking radiation is seeded by the non-equilibrium spatial distribution. It can be observed on the panel b) as density waves propagating inwards inside and outwards outside the horizon.

Hawking radiation in BECs has been largely studied in previous works (see e.g. \cite{Garay} for low-wavelength limit and \cite{Macher},\cite{Schutzhold} for the general case) and is not the main object of the present paper, which is why we restrain ourselves to the numerical demonstration of this effect, even though the possibility of experimental study of the Hawking radiation was one of the main reasons for the beginning of the activity on acoustic horizons\cite{Unruh}.

\emph{Wormholes} If a single scalar condensate is a model of a Universe which might contain some BHs, it is natural to map two spinor components to 2 different Universes. They can be completely decoupled from each other if there are no interactions between the particles of different spins. Adding a magnetic field can provide a coupling between these two Universes, making possible simulation of wormholes  \cite{Wheeler}.

In astrophysics, inter-Universe wormholes are the pairs of singularities located in different Universes and connected together. Using such wormhole, one can pass from one Universe to the other. A more interesting situation, when both holes connected together are in the same Universe, is called an intra-Universe wormhole. Such wormhole can connect two distant regions of space with a tunnel much shorter than the distance between the two, which might allow faster-than-light travel \cite{MorrisThorne}.

We start with the simpler case of an inter-Universe wormhole. We consider a 1D quantum wire, as in \cite{Wertz2010}. The idea is to first create a closed BH bordered by two event horizons and to connect the latter with a white hole in the other spin component using a local effective magnetic field $H_{eff}$, induced by the energy splitting existing between linearly polarized (TE or TM) eigen modes. $H_{eff}$ be controlled by varying the width of the wire \cite{Dasbach2005}, or  by applying an electric field \cite{APLMalpuech}.

In general relativity, a key concept is the propagation of signals, whose speed can never exceed that of light in the vacuum. The propagation of phonon wavepackets across the event horizons in BECs has already been studied, for example in refs. \cite{Carusotto2008,Mayoral2010}. In our model system, the "signals" will be grey solitons \cite{PitaevskiiBook} created in a single spin component \cite{VolovikHalfSoliton,FlayacHalfSoliton} (half-solitons) of the polariton condensate (e.g. $\sigma^{+}$, representing our Universe). Such solitons have a lot of properties similar to those of relativistic particles \cite{PitaevskiiBook,VolovikReview}, except that their mass is negative (because they are actually holes rather than particles): $m_{sol}=m_{0}/\sqrt{1-v^{2}/c_s^{2}}$; their size is given by $l_{sol}=\xi/\sqrt{1-v^{2}/c_s^{2}}$ where $v$ is their velocity, $c_s$ is the speed of sound and $\xi$ is the healing length of the condensate. For attractive interactions, the mass can of course be positive. We create pairs of  half-solitons thanks to a short time-dependent pulsed potential acting on a single spin component $U(x,t)=U_{0}\exp(-(x-x_0)^2/w_{x}^2)\exp(-(t-t_0)^2/\tau^2)$. As the speed of a soliton is related to its depth by $v=c_{s}n(0)/n$ where $n(0)$ is the density at its center and $n$ in its surrounding\cite{PitaevskiiBook}, we need a weak enough potential $U(x,t)$ to excite shallow solitons that will be able to travel at speeds close to $c_s$ in both spin components (marked as $c_1$,$c_2$ below).

\begin{figure}
  \includegraphics[width=0.4\textwidth]{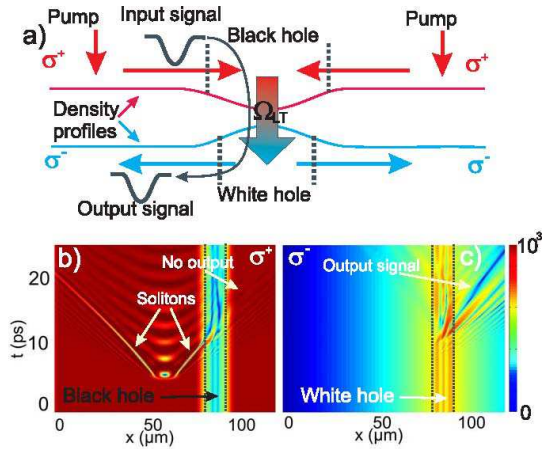}\\
  \caption{(Color online) a) Scheme of a wormhole between the $\sigma^+$ (red) and $\sigma^+$ (blue) Universes. Arrows show the directions of the flow in the two components. Dashed lines mark the event horizons in both components. b) Results of numerical simulation: $\left|\psi(x,t)\right|^2$ for both spin components with black dashed lines showing the boundaries of the black/white holes and white arrows indicating the propagation of signals. }
  \label{Fig3}
\end{figure}

Figure \ref{Fig3} shows the scheme of a numerical experiment with a single wormhole (panel (a)) and the results of the simulations (panels b,c). A 1D polariton wire is $cw$-pumped by two spatially separated, quasi-resonant, $\sigma^{+}$ polarized lasers allowing the formation of a steady state flow and specific density distribution  sketched on the panel (a). Polaritons flow away from the pumping regions and a closed BH is formed at the crossing of the flows. An effective magnetic field converting $\sigma^{+}$ to $\sigma^{-}$ is present in the BH region. The $\sigma^{-}$ density therefore shows a maximum expelling excitations outside from the central region which corresponds to the formation of a white hole in the $\sigma^{-}$ universe. After the steady state is obtained, a pulsed potential is applied in the $\sigma^{+}$-component (panel b) at $t=5$ ps at the left of the left horizon, creating a pair of propagating half-solitons. One of these propagates freely to the left, whereas the other enters the BH and remains partly guided inside. The effective magnetic field converts a part of this soliton into the other spin component ($\sigma^{-}$, panel c). The soliton is then able to cross the horizon of the white hole, propagating away together with the flowing condensate. One can also see that short-wavelength perturbations are still able to cross the horizon of the BH in any direction. The holes possess an internal structure, which we do not discuss here in details. The speed of the flow is zero at the center and the BH is in fact composed by two narrower BHs surrounding a subsonic region.

In the last part, we propose a scheme for an intra-Universe wormhole allowing for the transfer of a half-soliton with an apparent velocity faster than the speed of sound of its original universe.
Such intra-Universe wormhole is based on two inter-Universe wormholes similar to the the ones previously described, but connecting the Universes in opposite directions. The scheme (fig 4a) shows the proposed $\sigma^{\pm}$ density profiles with the two wormholes. Dashed line indicates the propagation of a half-soliton. A $\sigma^+$ half-soliton is generated at the left. It enters in the $\sigma^+$ BH where it is converted in a $\sigma^-$ soliton, which is ejected by the white hole part of the wormhole. Then, it travels in the $\sigma^-$ component between the two wormholes with a velocity close to $c_2$. It then reaches the second wormhole, which is a BH in the $\sigma^-$ component, where it is captured and converted to a $\sigma^+$ soliton ejected from the white hole. The average velocity of this soliton is close to $c_2$ which can be larger than $c_1$, the speed of sound in the $\sigma^+$ universe. The results of corresponding simulations are presented in panels (b) and (c). 

The pumping is $cw$ quasi-resonant with inhomogeneous elliptical polarization, providing the density profiles close to fig. 4(a). A pair of half-solitons is created at $t=5$ ps in the $\sigma^{+}$-component at $x=0$ (panel b). The "reference" half-soliton propagates to the left with the speed limited by $c_1$ and arrives to the edge at around $t=30$ ps. The half-soliton falling inside the BH converts into the $\sigma^{-}$ component, gets out of the white hole in $\sigma^{-}$ (panel c) and propagates with a higher speed, limited by $c_2>c_1$. This half-soliton arrives to the second wormhole and is converted back into the $\sigma^{+}$-component, appearing there at $t=25$ ps (marked by the green circle). The two events are marked with black horizontal lines, and the time difference between them is $\Delta t\approx5$ ps.

\begin{figure}
  \includegraphics[width=0.4\textwidth]{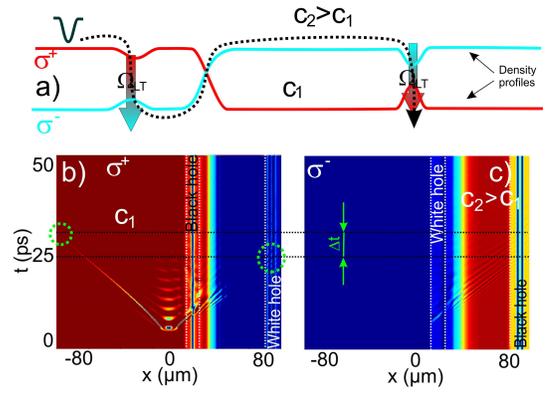}\\
  \caption{(Color online) "Faster-than-sound" signal propagation with two wormholes. a) Scheme with two wormholes showing density profiles; b) $\sigma^{+}$ density; c)$\sigma^{-}$ density. Dashed horizontal lines mark the arrival of the two signals (green dashed circles), with $\Delta t$ the time difference between them.}
  \label{Fig4}
\end{figure}

\emph{Conclusions} Spinor polariton condensates, being relatively easy to produce and manipulate, can be used for the simulation of astrophysical objects, such as black holes and wormholes. A 2D black hole with a closed event horizon can be simulated. Effective magnetic fields, well known as the cause of non-trivial spin dynamics of polaritons, can be used to organize the coupling between the black holes and white holes in the two spin components. A system of two wormholes in opposite directions allows one to organize "faster-than-sound" signal propagation.

\end{document}